\documentclass[a4paper,11pt]{article}
\usepackage{xcolor}
\usepackage{jheppub}
\usepackage{float,amssymb}
\usepackage{microtype}

\newcommand{\lambdabar}{{\mkern0.75mu\mathchar'26\mkern-9.75mu\lambda}}

\makeatletter
\gdef\@fpheader{}
\makeatother
\title{Gravitational spin Hall effect in the Reissner-Nordstr\"om black holes: evolution equations and charge effects}
	
\author{Zhan Liu,}

\author{Jia-Hui Huang}
\emailAdd{huangjh@m.scnu.edu.cn}
\affiliation{Key Laboratory of Atomic and Subatomic Structure and Quantum Control (Ministry of Education), Guangdong Basic Research Center of Excellence for Structure and Fundamental Interactions of Matter, School of Physics, South China Normal University, Guangzhou 510006, China}
\affiliation{Guangdong Provincial Key Laboratory of Quantum Engineering and Quantum Materials, Guangdong-Hong Kong Joint Laboratory of Quantum Matter, South China Normal University, Guangzhou 510006, China}
	
	\abstract{We study the spinoptics equations for test electromagnetic wave in the Reissner-Nordstr\"om spacetime. The explicit and hidden symmetries of this background allow the complex null tetrad parallel-propagated along the null geodesic to be constructed explicitly. From this tetrad we obtain the evolution equations for the deviation of the ray from the null geodesic and for the tilting angle of its orbital plane. In these equations the charge term enters with a sign opposite to the mass term. Numerical integration confirms that the gravitational spin Hall effect becomes weaker with increasing charge: at a fixed impact parameter away from the critical value, the magnitude of the dimensionless tilting angle decreases with $Q$. As the turning point approaches the  photon sphere, its magnitude grows firstly, and then exhibits an oscillatory behavior. Since the physical tilting angle is proportional to the reduced wavelength, it remains small whenever the spinoptics approximation is applicable. A divergence occurs only when the ray travels on the photon sphere.}
	
	
\begin{document}
		
		\maketitle
		\flushbottom
		
		\section{Introduction}
		\label{sec:1}
		
		In curved spacetime, the propagation of high-frequency electromagnetic waves is conventionally described by the geometric optics approximation\cite{Misner:1973prb, Dolan:2017zgu}. In this limit, a photon is treated as a point particle moving along a null geodesic, with its polarization vector parallel-propagated along the ray. However, for a realistic electromagnetic wave with a large but finite frequency, the spin-orbit coupling between the photon's spin and the spacetime curvature can no longer be neglected, which causes the actual propagation trajectory to acquire a subleading-order transverse shift depending on its polarization state, thereby deviating from the null geodesic. This phenomenon, which induces an orbital splitting of left- and right-handed circularly polarized light in a gravitational field, is known as the gravitational spin Hall effect (G-SHE). In astrophysical contexts, the G-SHE breaks the polarization degeneracy of ray trajectories in phenomena such as gravitational lensing and black hole scattering. 
		
		Current approaches to explaining the G-SHE of electromagnetic waves include generalizing the Mathisson-Papapetrou-Dixon equations, which describe the motion of massive spinning particles, to the massless case~\cite{Souriau:1974,Saturnini:1976hcd,Duval:2007eb,Duval:2016hxo,Duval:2018hzh}, applying the Foldy-Wouthuysen transformation to the Bargmann-Wigner equations~\cite{Gosselin:2006wp}, and employing modified geometric optics (or spinoptics)~\cite{Oancea:2020khc,Frolov:2020uhn,Frolov:2024ebe,Frolov:2024qow,Dahal:2022gop}. In the spinoptics framework, subleading-order terms in the high-frequency approximation are incorporated into the eikonal equation, so the worldline of a massless spinning particle is no longer a strict null geodesic but undergoes a deviation induced by the spin-curvature coupling. Harte and Oancea have shown that the gravitational spin Hall equations, which are equivalent at the subleading order to the spinoptics framework used here, arise as a special case of the MP equations~\cite{Harte:2022dpo}.
		
		The spinoptics method has been investigated in specific backgrounds\cite{Murk:2024qgj, Frolov:2024olb, Frolov:2012zn, Yoo:2012vv, Dahal:2023ncl, Frolov:2025bva}; it has recently been extended to parametrized and hairy spherically symmetric spacetimes~\cite{Alves:2026jyc} as well as axionic theories~\cite{Takeuchi:2026pyi}. In the Schwarzschild spacetime, Frolov used both explicit and hidden spacetime symmetries to construct a complex null tetrad parallel-propagated along the background null geodesic, showing that the primary consequence of the spin-curvature interaction is an asymptotic tilting of the massless particle's orbital plane, with the direction of the tilt depending on the particle's polarization helicity~\cite{Frolov:2024olb}. A similar construction using the principal tensor and the Carter constant was later carried out for the Kerr spacetime~\cite{Frolov:2025bva}, where it was shown that the black hole spin modifies the asymptotic tilting angle, with the modification depending on whether the ray is prograde or retrograde.
		
		To explore the G-SHE beyond Ricci-flat geometries, we extend spinoptics analysis to the Reissner-Nordstr\"om (RN) spacetime. From an astrophysical perspective, previous studies demonstrated that realistic black holes can sustain a non-zero, albeit negligibly small, electric charge\cite{Wald:1974np, Bally:1978, Zajacek:2018ycb}. Recent studies indicate that primordial black holes may carry charges under an additional $U(1)$ gauge symmetry in the dark sector~\cite{Bai:2019zcd, Baker:2025cff}. If the corresponding dark electron is much heavier than the black hole's Hawking temperature, dark electrons are not emitted as Hawking radiation and the dark charge does not change. The black hole instead loses mass via Hawking radiation of massless particles and becomes quasiextremal, a state in which the Hawking temperature is strongly suppressed. The Schwinger discharge, inhibited by the heavy dark electron, becomes efficient only in the late stages of evaporation, so the black hole can maintain a high charge-to-mass ratio over cosmological timescales. Such primordial near-extremal black holes (PeBHs) have been proposed as candidates for dark matter and as sources of the PeV neutrinos. Their exterior geometry is described by the RN metric, which provides a motivation for studying charge-dependent effects in polarized-wave propagation.
		
		In this work, following the spinoptics framework established for the Schwarzschild background~\cite{Frolov:2024olb}, we study the deviation of the ray from the null geodesic, the tilting angle of its orbital plane, and their dependence on the black hole charge. In Sec.~\ref{sec:2}, we outline the transition from geometric optics to spinoptics, introduce the closed conformal Killing-Yano tensor of the RN spacetime, and review the dynamics of null geodesics together with the location of the photon sphere. In Sec.~\ref{sec:3}, we construct the parallel-propagated null tetrad using the hidden symmetry, compute the driving force $w^\mu$, and give the evolution equations for the rescaled deviation $\xi\equiv r\Psi$ and the tilting angle $\vartheta$. Sec.~\ref{sec:4} presents the numerical results for the tilting of the orbital plane and examines how the black hole charge alters the G-SHE. We conclude in Sec.~\ref{sec:5}.

		In this paper we use geometric units with $c = G = 1$ and adopt the metric signature $(-,+,+,+)$. Latin indices $a,b,c,\ldots$ denote abstract indices and refer to tensors themselves, while Greek indices $\mu,\nu,\rho,\ldots$ denote their coordinate components.
		
		\section{Theoretical Framework}
		\label{sec:2}
		
		\subsection{From Geometric Optics to Spinoptics}
		
		In curved spacetime, the propagation of electromagnetic waves is governed by the source-free Maxwell equations $\nabla^b F_{ab} = 0$, $\nabla_{[a} F_{bc]} = 0$. In astrophysical contexts, the wavelength of electromagnetic waves is typically much smaller than the spacetime curvature radius, a regime in which the geometric optics approximation provides a natural description. Under this approximation, the complex vector potential is assumed to take an asymptotic expansion ansatz containing a rapidly varying phase:
		\begin{equation}
			\mathcal{A}_a = A_a e^{iS},
		\end{equation}
		where $\nabla_a S = \omega p_a$. In the high-frequency limit ($\omega \to \infty$), the Maxwell equations are truncated at the leading order, yielding the classical eikonal equation $p^a p_a = 0$, while requiring the polarization vector to be strictly parallel-propagated along the ray. In this limiting case, the photon is treated as a geometric point evolving along a null geodesic, completely decoupling its intrinsic spin degrees of freedom from the background curvature.
		
		However, realistic electromagnetic waves carry spin, and their transverse polarization planes are inevitably affected by the geometric structure of spacetime (i.e., the polarization vector is no longer strictly parallel-propagated). To capture this effect along the ray trajectory, a ray theory preserving wave characteristics, known as spinoptics, was developed\cite{Oancea:2020khc, Frolov:2020uhn}.
		
		To isolate the effects of different spins, one can utilize the Hodge dual and the real electromagnetic field tensor $F_{ab}$ to construct the complex electromagnetic field tensor corresponding to different helicities:
		\begin{equation}
			(\mathcal{F}^s)_{ab} = \frac{1}{2} (F_{ab} - i s {}^*F_{ab}).
		\end{equation}
		
		By definition, this construction strictly restricts the right-handed ($s=+1$) and left-handed ($s=-1$) polarization states to self-dual and anti-self-dual solutions, respectively.
		
		Practically, we describe this field by introducing the ansatz $(\mathcal{A}^s)_a = (A^s)_a e^{iS^s}$. The complex electromagnetic field tensor directly generated by this ansatz is given by ${\tilde{\mathcal{F}}}^s = \mathrm{d}{\mathcal{A}}^s$. However, $(\tilde{\mathcal{F}}^s)_{ab}$ is not \textit{a priori} (anti-)self-dual. To ensure that $(\tilde{\mathcal{F}}^s)_{ab}$ correctly represents the physical pure-helicity state $(\mathcal{F}^s)_{ab}$, we must explicitly impose the (anti-)self-duality condition on $(\tilde{\mathcal{F}}^s)_{ab}$\cite{Frolov:2020uhn}. This constraint ensures that during evolution, a right-handed (or left-handed) polarized wave must maintain its purely (anti-)self-dual nature without mixing with opposite states, requiring the corresponding expansion coefficients of the opposite helicity to vanish.
		
		In spinoptics, one does not solve the asymptotic expansion order by order in isolation\cite{Misner:1973prb}. Instead, to capture the coupling between spin and spacetime geometry, the subleading polarization effects must be incorporated into the leading-order dynamics. At this level of precision, we can construct an effective Hamiltonian determining the phase-space trajectory of the high-frequency electromagnetic waves\cite{Frolov:2020uhn,Frolov:2024ebe}:
		\begin{equation}
			H = \frac{1}{2\omega} (P^\mu - s b^\mu)(P_\mu - s b_\mu),
		\end{equation}
		where $b_\mu = i \bar{m}^\nu \nabla_\mu m_\nu$ and $P_\mu=\omega p_\mu$ is the canonical momentum, which modifies the photon's dispersion relation. At this stage, the spin-orbit coupling is fully restored within the dynamical framework. 
		
		Structurally, this effective Hamiltonian is formally analogous to that of a charged particle in an electromagnetic field. In this analogy, $b_a$ acts as an effective $U(1)$ gauge potential, while the helicity $s$ serves as the effective charge. Consequently, rays with opposite helicities undergo transverse deviations in opposite directions. This polarization-dependent orbital splitting is known as the gravitational spin Hall effect.
		
		To apply this framework to a concrete spacetime, one needs the Riemann curvature components and a parallel-propagated null tetrad. In the following, we introduce the hidden symmetry of the RN spacetime that makes the analytic construction of this tetrad possible.
		
		\subsection{Closed Conformal Killing-Yano Tensor of the RN Spacetime}
		
		For the RN black hole, we adopt the coordinates $(t, r, \theta, \phi)$ adapted to its static and spherically symmetric geometric properties. The line element is given by $ds^2 = -f dt^2 + f^{-1} dr^2 + r^2 (d\theta^2 + \sin^2\theta d\phi^2)$, where the metric function $f = 1 - 2M/r + Q^2/r^2$ incorporates the black hole mass $M$ and charge $Q$.
		
		As a Petrov type D spacetime, the RN metric can be recovered from the Carter metric and possesses a hidden symmetry generated by the second-rank closed conformal Killing-Yano tensor (CCKYT) $h_{ab}$\cite{Frolov:2017kze}. The Hodge dual of this antisymmetric 2-form field is the Killing-Yano tensor $k_{ab} = \frac{1}{2} h^{cd} \epsilon_{abcd}$. By utilizing $h_{ab}$ and $k_{ab}$, we can directly construct a complex null tetrad $(l^a, m^a, \bar{m}^a, n^a)$ that is parallel-propagated along the geodesics from the null vector $l^a$, which we use below.
		
		The CCKYT in the RN spacetime takes the same coordinate form as in Schwarzschild,
		$h_{\mu\nu} = r(\delta_\mu^t \delta_\nu^r - \delta_\mu^r \delta_\nu^t)$,
		which can be derived from the potential $\beta_a = -\frac{1}{2}r^2 (dt)_a$ via $h_{ab} = 2\nabla_{[a} \beta_{b]}$, and its Hodge dual gives the Killing-Yano tensor.
		When raising indices to compute $h^{ab}$, the metric function $f(r)$ cancels exactly because $g^{tt}g^{rr} = (-1/f) \cdot f = -1$,
		and $\sqrt{-g} = r^2\sin\theta$ is also independent of $f(r)$. Consequently, both $h_{ab}$ and $k_{ab}$ take the same coordinate form as in the Schwarzschild case.

		\subsection{Dynamics of Null Geodesics and the Photon Sphere}
		
		The RN spacetime admits explicit symmetries generated by the timelike Killing vector field $\xi^a = (\partial/\partial t)^a$ and three rotational Killing vector fields $\zeta_{(i)}^a$ ($i=1,2,3$). These geometric symmetries imply the existence of the conserved energy $E = -\xi^a l_a$ and the axial angular momentum $L_z = \zeta_{(1)}^a l_a$ (where $\zeta_{(1)}^a = (\partial/\partial \phi)^a$). In addition, the total angular momentum square can be expressed as a constant of motion $L^2 = K_{ab} l^a l^b$. Geometrically, $K_{ab}$ is a symmetric second-rank Killing tensor, which is reducible in the spherically symmetric RN spacetime. Although $L^2$ can be fully derived from the explicit rotational Killing vectors, the Killing tensor can also be expressed via the Killing-Yano tensor as
		\begin{equation}
			K_{ab} = k_a^{\ c} k_{cb}.\label{eq:KfromKY}
		\end{equation}
		In the spherically symmetric RN spacetime the two constructions are equivalent; \eqref{eq:KfromKY} has the advantage that it generalizes directly to the charged Kerr–NUT–(A)dS family of spacetimes, where the Killing tensor becomes irreducible and the Killing-Yano tensor provides its natural geometric origin. Ultimately, with the constants of motion $E$, $L_z$, and $L^2$ completely specified, the tangent vector $l^a$ of the background null geodesic can be explicitly determined:
		
		\begin{equation}
			l^\mu = \left( \frac{E}{f}, \frac{\mathcal{R}}{r}, \frac{\Theta}{r^2 \sin\theta}, \frac{L_z}{r^2 \sin^2\theta} \right),
		\end{equation}
		where the energy $E$ and the azimuthal angular momentum $L_z$ are constants of motion, and we have:
		\begin{equation}
			\mathcal{R} = \pm (E^2 r^2 - L^2 f)^{1/2}, \quad \Theta = \pm (L^2 \sin^2\theta - L_z^2)^{1/2}.
		\end{equation}
		
		We can always select a specialized coordinate system such that the null geodesic lies in the equatorial plane ($\theta = \pi/2$), where $\Theta = 0$ and $L_z = L$. By further choosing an appropriate affine parameter to normalize the energy to $E=1$, the impact parameter becomes $l \equiv L/E = L$. The components of the null geodesic tangent vector then simplify to:
		\begin{equation}
			l^\mu = \left( \frac{1}{f}, \frac{\mathcal{R}}{r}, 0, \frac{L}{r^2} \right), \quad \mathcal{R} = \pm (r^2 - L^2 f)^{1/2}.
		\end{equation}
		
		From the radial equation of the null geodesic $(dr/d\lambda)^2 + V_{\text{eff}} = 1$, where the effective potential is $V_{\text{eff}} = \frac{L^2}{r^2} (1 - \frac{2M}{r} + \frac{Q^2}{r^2})$, the photon sphere of the RN spacetime is located at~\cite{Chandrasekhar:1983,Chen:2022ewe,Huang:2025qsi}:
		\begin{equation}
			r_{\rm ph} = \frac{1}{2} \left[ 3M + (9M^2 - 8Q^2)^{1/2} \right].
		\end{equation}
		In the subsequent scattering problem analysis, the radial turning point $r_m$ of the actual ray must lie outside this photon sphere.
		
		\section{Deviation of the Actual Ray from the Null Geodesic}
		\label{sec:3}
		
		According to the spinoptics framework, the evolution of the actual ray $\gamma(\lambda)$ satisfies the modified geodesic equation\cite{Frolov:2020uhn}:
		\begin{equation}
			l^b \nabla_b l^a = w^a,
			\label{eq:spinoptics}
		\end{equation}
		where $\lambda$ is the affine parameter, and $l^a$ is the ray's tangent vector. The spin-curvature coupling term $w^a$ on the right-hand side is determined by the Riemann curvature $R_{abcd}$ and the wave's transverse polarization plane. For a circularly polarized wave with helicity $s = \pm 1$, this term is explicitly expressed by the null tetrad as:
		\begin{equation}
			w^a = \frac{s}{\omega} (\bar{\kappa} m^a + \kappa \bar{m}^a),
		\end{equation}
		where $\kappa = i R_{abcd} l^a m^b m^c \bar{m}^d$. Here, the complex null tetrad basis vectors $m^a$ and $\bar{m}^a$ span the polarization plane. Strictly speaking, the null tetrad appearing in the term $w^a$ is defined on the actual ray trajectory $\gamma(\lambda)$ and must be F-transported along it \cite{Frolov:2020uhn}. However, the components $w^\mu$ are of order $\mathcal{O}(\omega^{-1})$. Evaluating $w^\mu$ on a nearby null geodesic $\gamma_0(\lambda)$---rather than on $\gamma(\lambda)$ itself---introduces only $\mathcal{O}(\omega^{-2})$ corrections, which lie beyond our approximation. We may therefore evaluate $w^a$ using the parallel-propagated tetrad $((l_0)^a, (n_0)^a, (m_0)^a, (\bar{m}_0)^a)$ constructed on $\gamma_0$, where the F-transported condition reduces to ordinary parallel transport. In general spacetimes, solving for this parallel-propagated tetrad typically relies on tedious numerical integration. However, in spacetimes of specific algebraic types, we can analytically construct it by utilizing hidden symmetries.
		
		Using the $h_{ab}$ and $k_{ab}$, we first construct a set of vectors~\cite{MARCK1983140,03e6e95f-6105-3f53-9aa8-4864f05e2d86,Connell:2008vn} associated with $(l_0)^a$ satisfying orthonormal conditions in the $(t, r, \theta, \phi)$ coordinates:
		\begin{equation}
			(\tilde{e}_1)^\mu = \frac{1}{L} h^\mu_{\ \nu} (l_0)^\nu = \left( \frac{\mathcal{R}}{Lf}, \frac{r}{L}, 0, 0 \right),
		\end{equation}
		\begin{equation}
			(e_2)^\mu = \frac{1}{L} k^\mu_{\ \nu} (l_0)^\nu = \left( 0, 0, \frac{1}{r}, 0 \right),
		\end{equation}
		\begin{equation}
			(\tilde{e}_3)^\mu = \frac{r^2}{2L^2} \left( \frac{1}{f}, \frac{\mathcal{R}}{r}, 0, -\frac{L}{r^2} \right).
		\end{equation}
		Further, by using $((l_0)^a, (\tilde{e}_1)^a, (e_2)^a, (\tilde{e}_3)^a)$ the parallel-propagated tetrad associated with the null geodesic can be constructed as:
		\begin{align}
			(l_0)^\mu &= \left( \frac{1}{f}, \frac{\mathcal{R}}{r}, 0, \frac{L}{r^2} \right), \\
			(n_0)^a &= (\tilde{e}_3)^a - \Phi (e_1)^a - \frac{1}{2} \Phi^2 (l_0)^a, \\
			(m_0)^a &= \frac{1}{\sqrt{2}} \left[ (e_1)^a + i(e_2)^a \right], \\
			(\bar{m}_0)^a &= \frac{1}{\sqrt{2}} \left[ (e_1)^a - i(e_2)^a \right],
		\end{align}
		where $(e_1)^a = (\tilde{e}_1)^a - \Phi (l_0)^a$ is parallel-transported, and $\Phi$ satisfies $d\Phi/d\lambda = E/L$.
		
		Evaluating $w^a$ with this null tetrad reveals that the perturbation term has a non-zero component only in the polar ($\theta$) direction:
		\begin{equation}
			w^\mu = \left( 0, 0, \frac{s}{\omega} \frac{l^2 \Phi}{r^7} (2Q^2 - 3Mr), 0 \right)\label{eq:wcomp}.
		\end{equation}

		Because $w^a$ of~\eqref{eq:wcomp} points along the polar direction $(e_2)^a$, this transverse force drives the actual ray away from the original geodesic. We parametrize both $\gamma(\lambda)$ and $\gamma_0(\lambda)$ with the affine parameter $\lambda$ of $\gamma_0$ and define
		\begin{equation}
			\delta x^\mu \equiv \left. x^\mu \right|_{\gamma(\lambda)} - \left. x_0^\mu \right|_{\gamma_0(\lambda)}. \label{eq:deltax}
		\end{equation}
		Written in coordinates along $\gamma(\lambda)$, the spinoptics equation~\eqref{eq:spinoptics} reads
		\begin{equation}
			\left. w^\mu \right|_{\gamma(\lambda)} = \left( \frac{d^2x^\mu}{d\lambda^2} + \Gamma^\mu_{\ \nu\sigma}\frac{dx^\nu}{d\lambda}\frac{dx^\sigma}{d\lambda} \right) \bigg|_{\gamma(\lambda)},
			\label{eq:spinopticse}
		\end{equation}
		while the geodesic $\gamma_0$ satisfies the same equation with $w^\mu=0$. Subtracting the latter equation from Eq.\eqref{eq:spinopticse}, and using the multivariate Taylor expansion of the Christoffel symbols
		\begin{equation}
			\Gamma^\mu_{\ \nu\sigma}(x_0+\delta x) = \Gamma^\mu_{\ \nu\sigma}(x_0) + \delta x^\rho\, \partial_\rho\Gamma^\mu_{\ \nu\sigma}(x_0) + \mathcal{O}\left[(\delta x)^2\right],
		\end{equation}
		we obtain, to first order in $\delta x^\mu$,
		\begin{equation}
			\frac{d^2\delta x^\mu}{d\lambda^2} + 2\left. \Gamma^\mu_{\ \nu\sigma} \right|_{\gamma(\lambda)} \frac{dx_0^\nu}{d\lambda}\frac{d\delta x^\sigma}{d\lambda} + \frac{dx_0^\nu}{d\lambda}\frac{dx_0^\sigma}{d\lambda} \left. \partial_\rho\Gamma^\mu_{\ \nu\sigma}(x_0) \right|_{\gamma_0(\lambda)} \delta x^\rho = \left. w^\mu \right|_{\gamma(\lambda)}. \label{eq:deviation}
		\end{equation}
		
		We choose the turning point to be $\lambda=0$, where the two rays coincide both in position and in tangent direction,
		\begin{equation}
			\delta x^\mu(0) = 0, \qquad \left. \frac{d\delta x^\mu}{d\lambda} \right|_{\lambda=0} = 0. \label{eq:IC}
		\end{equation}
		With this initial data, the $t$, $r$ and $\phi$ components of $\delta x^\mu$ vanish at order $\mathcal{O}(\omega^{-1})$, so only the polar component survives
		\begin{equation}
			\delta x^\mu = \left( 0, 0, \vartheta, 0 \right), \qquad \vartheta \equiv \delta x^\theta, \label{eq:thetaonly}
		\end{equation}
		as explained in Appendix~\ref{app:deviation}. The $\theta$ component of~\eqref{eq:deviation} gives
		\begin{equation}
			\frac{d^2\vartheta}{d\lambda^2} + \frac{2}{r}\frac{dr}{d\lambda}\frac{d\vartheta}{d\lambda} + \frac{l^2}{r^4}\vartheta = w^\theta. \label{eq:thetaeq}
		\end{equation}

		Following the notation of Ref.~\cite{Frolov:2024olb}, the polar component of the ray's tangent vector is written as
		\begin{equation}
			\left. l^\theta \right|_{\gamma(\lambda)} = \left. l_0^\theta \right|_{\gamma_0(\lambda)} + \frac{\Psi}{r}, \qquad \Psi \equiv  r\frac{d\vartheta}{d\lambda}, \label{eq:ltheta}
		\end{equation}
		so that the tangent vectors of the two rays differ only through $\Psi$. 		Introducing $\xi \equiv r\Psi = r^2\, d\vartheta/d\lambda$ and multiplying~\eqref{eq:thetaeq} by $r^2$, we obtain
		\begin{equation}
			\frac{d\xi}{d\lambda} = r^2 w^\theta - \frac{l^2}{r^2}\vartheta, \qquad \frac{d\vartheta}{d\lambda} = \frac{\xi}{r^2}. \label{eq:xi_evo}
		\end{equation}
		Substituting $w^\theta$ from~\eqref{eq:wcomp} and using $\Phi=\lambda/l$, the first equation in Eqs.~\eqref{eq:xi_evo} takes the form
		\begin{equation}
			\frac{d\xi}{d\lambda} = s\lambdabar\,\frac{l\lambda}{r^5}\,(2Q^2-3Mr) - \frac{l^2}{r^2}\vartheta, \label{eq:xi_evo2}
		\end{equation}
		where $\lambdabar \equiv 1/\omega$.
		
		Introducing the dimensionless variables $\tilde{r} \equiv r/2M$, $\tilde{\lambda} \equiv \lambda/2M$, $\tilde{l} \equiv l/2M$, $\tilde{Q} \equiv Q/2M$, $\tilde{\lambdabar} \equiv 1/(2M\omega)$, $\tilde{\Psi} \equiv \Psi/(s\tilde{\lambdabar})$, $\tilde{\xi} \equiv \tilde{r}\tilde{\Psi}$ and $\tilde{\vartheta} \equiv \vartheta/(s\tilde{\lambdabar})$, Eqs.~\eqref{eq:xi_evo} becomes
		\begin{equation}
			\frac{d\tilde{\xi}}{d\tilde{\lambda}} = \left( 2\tilde{Q}^2 - \frac{3}{2}\tilde{r} \right)\frac{\tilde{l}\tilde{\lambda}}{\tilde{r}^5} - \frac{\tilde{l}^2}{\tilde{r}^2}\tilde{\vartheta}, \qquad \frac{d\tilde{\vartheta}}{d\tilde{\lambda}} = \frac{\tilde{\xi}}{\tilde{r}^2}, \label{eq:dxi_tilde}
		\end{equation}
		which will be studied numerically in the next section.

		\section{Numerical Calculation}
		\label{sec:4}
		
		[0]The deviation is obtained by integrating the coupled radial equation and evolution equations. The radial equation of the null geodesic follows from the first-order equation
		\begin{equation}
			\frac{dr}{d\lambda} = \pm \frac{1}{r} \bigl( r^2 - l^2 f \bigr)^{1/2}.
		\end{equation}
		This equation is singular at the turning point where $dr/d\lambda = 0$. Using the standard approach, we differentiate it to obtain a second-order form:
		\begin{equation}
			\frac{d^2\tilde{r}}{d\tilde{\lambda}^2} = \frac{\tilde{l}^2}{\tilde{r}^3} \left( 1 - \frac{3}{2\tilde{r}} + \frac{2\tilde{Q}^2}{\tilde{r}^2} \right).
		\end{equation}
		Together with the evolution equations~\eqref{eq:dxi_tilde}, these form a system of equations that we integrate numerically. For a given turning point $\tilde{r}_m$, the impact parameter is determined by the condition $d\tilde{r}/d\tilde{\lambda}=0$ at $\tilde{r}=\tilde{r}_m$, that is
		\begin{equation}
			\tilde{l} = \frac{\tilde{r}_m^2}{\sqrt{\tilde{r}_m^2 - \tilde{r}_m + \tilde{Q}^2}}.
			\label{eq:lOfRm}
		\end{equation}
		We take the affine parameter to vanish at the turning point and impose the initial conditions
		\begin{equation}
			\tilde{r}(0) = \tilde{r}_m, \qquad \left.\frac{d\tilde{r}}{d\tilde{\lambda}}\right|_{\tilde{\lambda}=0} = 0, \qquad \tilde{\vartheta}(0) = 0, \qquad \tilde{\xi}(0) = 0,
			\label{eq:numIC}
		\end{equation}
		the last two of which are the dimensionless form of Eq.~\eqref{eq:IC}.
		
		[1]First, we investigate the evolutionary behavior of the dimensionless deviation $\tilde{\Psi} = \tilde{\xi}/\tilde{r}$, which characterizes the polar component of the difference between the tangent vectors of the actual ray and of the null geodesic, during a single scattering process.
		\begin{figure}[htbp]
			\centering
			\begin{minipage}[t]{0.49\linewidth}
				\centering
				\includegraphics[width=\linewidth]{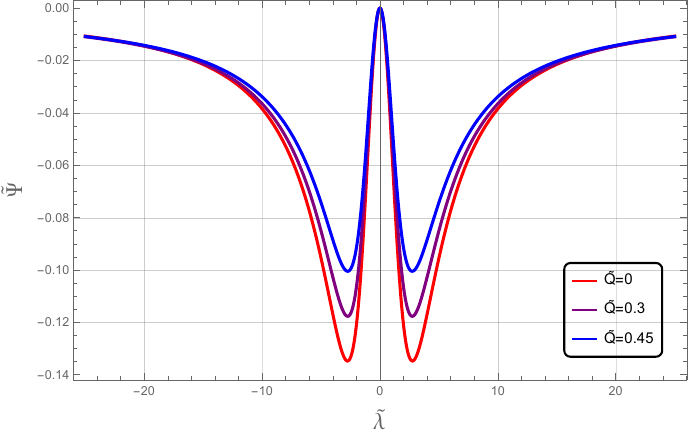}
			\end{minipage}\hfill
			\begin{minipage}[t]{0.49\linewidth}
				\centering
				\includegraphics[width=\linewidth]{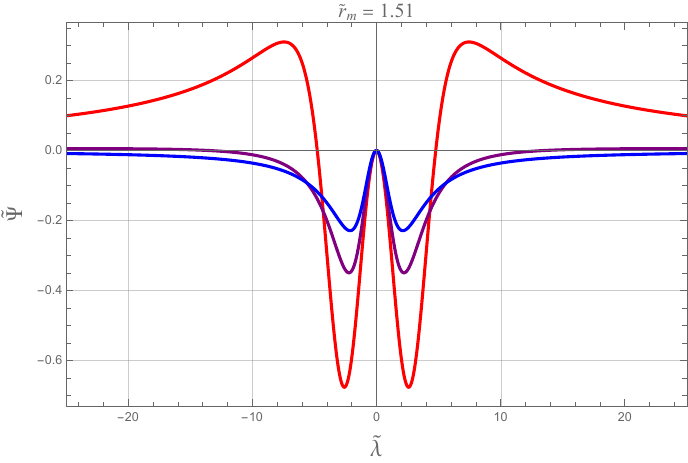}
			\end{minipage}
			\caption{Evolution of the dimensionless deviation $\tilde{\Psi}$ with respect to the affine parameter $\tilde{\lambda}$ for three charge values. (a) Left panel: $\tilde{r}_m=2.0$; (b) Right panel: $\tilde{r}_m=1.51$. The color coding is the same in both panels.}
			\label{fig:1}
		\end{figure}
		For definiteness, we consider two turning-point radii, $\tilde{r}_m=2.0$ and $\tilde{r}_m=1.51$, both lying outside the photon sphere for all three charge values $\tilde{Q}=0$ (Schwarzschild), $\tilde{Q}=0.3$, and $\tilde{Q}=0.45$ (near-extremal). For different $\tilde{Q}$, fixing $\tilde{r}_m$ means that rays have different impact parameters, according to Eq.~\eqref{eq:lOfRm}. This comparison at a fixed turning-point radius is similar to that adopted for the Kerr case in~\cite{Frolov:2025bva}.
		
		Figure~\ref{fig:1}(a) shows that the deviation accumulates in the same direction during both the infall ($\tilde{\lambda}<0$, from infinity to the turning point) and escape ($\tilde{\lambda}>0$, from the turning point to infinity) phases, giving an even-function profile. The magnitude of the deviation $|\tilde{\Psi}|$ first grows and then decays to zero as $\tilde{r}\to\infty$, since $\tilde{\Psi}_\infty = \tilde{\xi}_\infty/\tilde{r}_\infty$, where the convergence of $\tilde{\xi}_\infty$ can be verified. The maximum magnitude of the deviation decreases with increasing charge, reflecting the reduced accumulation rate.
		
		Figure~\ref{fig:1}(b) shows the same three charge values for $\tilde{r}_m=1.51$. This turning point is chosen so that the ray with $\tilde{Q}=0$ passes just outside the photon sphere. For $\tilde{Q}=0$ the deviation behaves differently: it first grows in the negative direction, then returns to zero, changes sign and grows in the opposite direction before decaying to zero again. For the larger charges the reversal is weaker or even absent.
		
		[2]We next consider the tilting angle of the ray with respect to the equatorial plane. Writing the polar angle of the actual ray as $\theta(\lambda) = \pi/2 + \vartheta(\lambda)$, the accumulated deflection is described by the dimensionless tilting angle $\tilde{\vartheta} = \vartheta/(s\tilde{\lambdabar})$ introduced in Sec.~\ref{sec:3}.
		\begin{figure}[htbp]
			\centering
			\begin{minipage}[t]{0.49\linewidth}
				\centering
				\includegraphics[width=\linewidth]{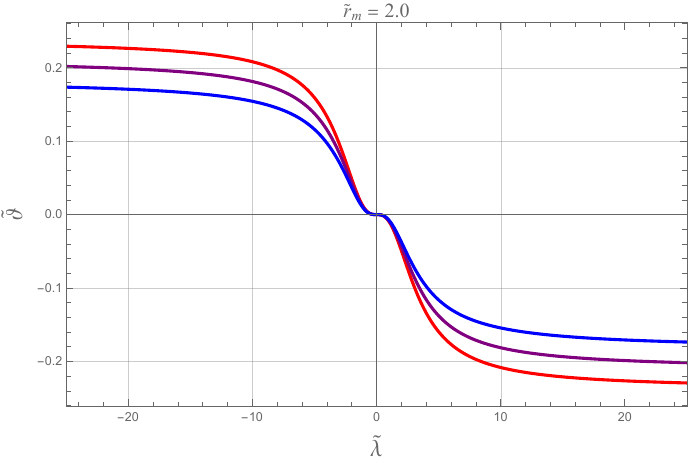}
			\end{minipage}\hfill
			\begin{minipage}[t]{0.49\linewidth}
				\centering
				\includegraphics[width=\linewidth]{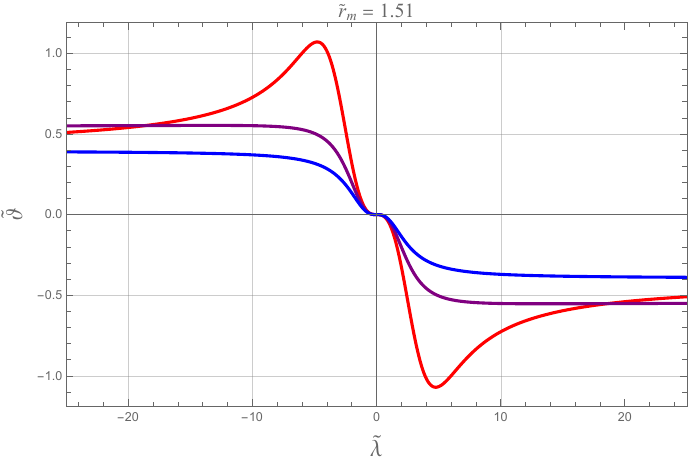}
			\end{minipage}
			\caption{Evolution of the dimensionless tilting angle $\tilde{\vartheta}$ with respect to the affine parameter $\tilde{\lambda}$ for three charge values, with the same color coding as Fig.~\ref{fig:1}. (a) Left panel: $\tilde{r}_m=2.0$; (b) Right panel: $\tilde{r}_m=1.51$.}
			\label{fig:2}
		\end{figure}
		Figure~\ref{fig:2} shows $\tilde{\vartheta}$ for cases with the same parameters as in Fig.~\ref{fig:1}. For $\tilde{r}_m=2.0$ (Fig.~\ref{fig:2}(a)) the magnitude of the tilting angle increases monotonically with $|\tilde{\lambda}|$, approaching its asymptotic value $\tilde{\vartheta}_\infty$. For $\tilde{r}_m=1.51$ (Fig.~\ref{fig:2}(b)), the accumulation for $\tilde{Q}=0$ is no longer monotonic: the tilting angle grows beyond its asymptotic value and then returns towards it. Since $\tilde{\Psi} = \tilde{r}\,d\tilde{\vartheta}/d\tilde{\lambda}$, the sign change of $\tilde{\Psi}$ in Fig.~\ref{fig:1}(b) corresponds to an extremum of $\tilde{\vartheta}$ in Fig.~\ref{fig:2}(b).
		
		Astrophysical observations are typically conducted in asymptotically flat regions far from the gravitational source. The quantity of physical interest is therefore the asymptotic value $\tilde{\vartheta}_\infty$, which characterizes the angular deflection accumulated over the entire scattering process. In the following, we study it as a function of the turning-point radius and of the impact parameter for different charge values.

\begin{figure}[htbp]
			\centering
			\begin{minipage}[t]{0.49\linewidth}
				\centering
				\includegraphics[width=\linewidth]{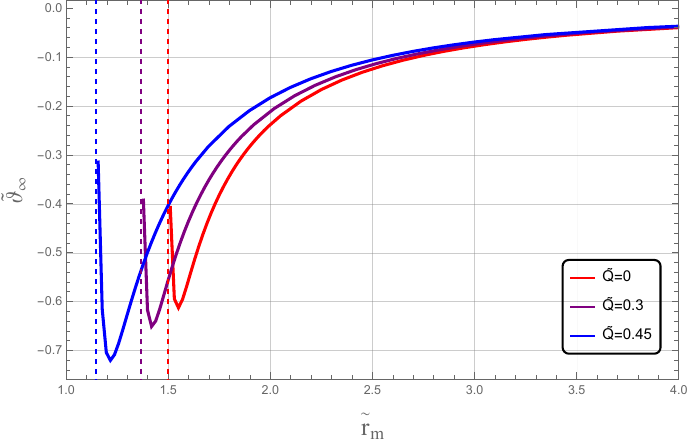}
			\end{minipage}\hfill
			\begin{minipage}[t]{0.49\linewidth}
				\centering
				\includegraphics[width=\linewidth]{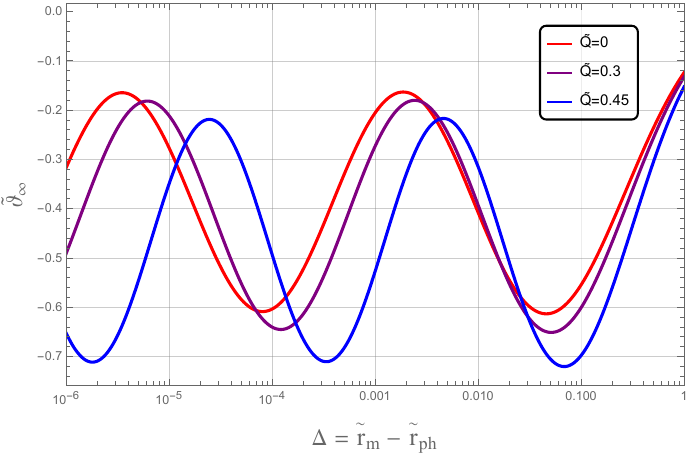}
			\end{minipage}
			\caption{Asymptotic tilting angle $\tilde{\vartheta}_\infty$ for three charge values. (a) Dependence on the turning-point radius $\tilde{r}_m$. (b) Dependence on $\Delta=\tilde{r}_m-\tilde{r}_{\rm ph}(\tilde{Q})$ near the photon sphere, with a logarithmic horizontal axis.}
			\label{fig:3}
		\end{figure}

		[3]Figure~\ref{fig:3}(a) shows $\tilde{\vartheta}_\infty$ as a function of $\tilde{r}_m$ for the three charge values, with the vertical lines indicating the corresponding photon spheres $\tilde{r}_{\rm ph}(\tilde{Q})$. The magnitude $|\tilde{\vartheta}_\infty|$ grows as $\tilde{r}_m$ decreases towards the photon sphere from spatial infinity. However, this growth does not continue when $\tilde{r}_m$ is close to the photon sphere: Fig.~\ref{fig:3}(b) shows $|\tilde{\vartheta}_\infty|$ as a function of $\Delta \equiv \tilde{r}_m - \tilde{r}_{\rm ph}(\tilde{Q})$ with a logarithmic horizontal axis, and the values remain bounded, exhibiting an oscillatory dependence on $\Delta$. The divergence occurs only on the exactly critical orbit ($\Delta=0$), where the ray remains on the circular photon orbit and the tilting angle grows linearly with the affine parameter (Appendix~\ref{app:critical}); for any $\Delta>0$ the ray leaves the vicinity of the photon sphere after a finite time.
		
\begin{figure}[htbp]
			\centering
			\includegraphics[width=0.6\linewidth]{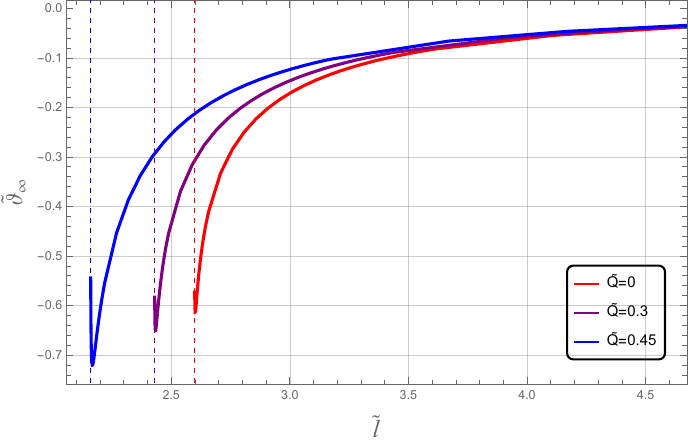}
			\caption{Asymptotic tilting angle $\tilde{\vartheta}_\infty$ as a function of the impact parameter for the three charge values.}
			\label{fig:4}
		\end{figure}

		The physical tilting angle accumulated at infinity is $|\vartheta_\infty| = \tilde{\lambdabar}\,|\tilde{\vartheta}_\infty|$, with $\tilde{\vartheta}_\infty$ of order unity. The deviation is therefore controlled by the same small parameter $\tilde{\lambdabar}$ that underlies the spinoptics approximation, and it remains small whenever this approximation is applicable\footnote{The spinoptics framework is built on a high-frequency expansion truncated at $\mathcal{O}(\omega^{-1})$, whose validity requires the reduced wavelength $\lambdabar=1/\omega$ to be much smaller than both the characteristic scale of variation of the electromagnetic field and the curvature radius of the spacetime~\cite{Frolov:2024ebe}.}, confirming that the trajectory stays close to the equatorial plane and that Eq.~\eqref{eq:deltax} is self-consistent. [4]We also compare rays with the same impact parameter. Figure~\ref{fig:4} shows $\tilde{\vartheta}_\infty$ as a function of $\tilde{l}$ for the same three charge values. The magnitude $|\tilde{\vartheta}_\infty|$ grows as $\tilde{l}$ approaches the critical impact parameter $\tilde{l}_{\rm crit}(\tilde{Q})$ (indicated by the vertical lines) corresponding to the photon sphere; the sharp rise near $\tilde{l}_{\rm crit}$ reflects the same bounded near-critical behavior examined in Fig.~\ref{fig:3}(b). For a fixed $\tilde{l}$ away from the critical value, a larger $\tilde{Q}$ yields a smaller $|\tilde{\vartheta}_\infty|$.
		
\begin{figure}[htbp]
			\centering
			\includegraphics[width=0.6\linewidth]{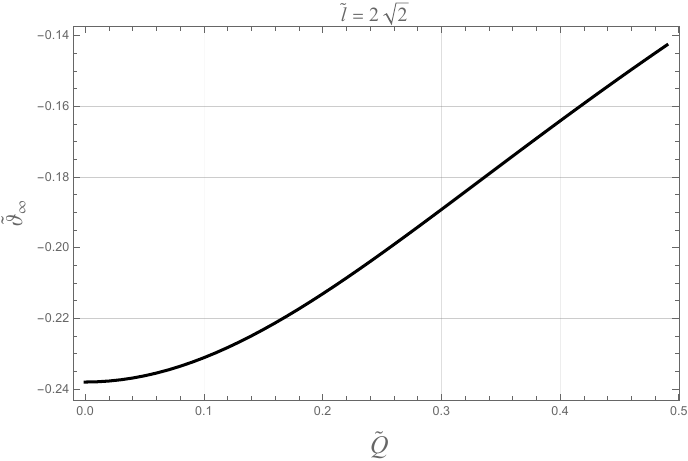}
			\caption{Asymptotic tilting angle $\tilde{\vartheta}_\infty$ as a function of $\tilde{Q}$ at a fixed impact parameter $\tilde{l} = 2\sqrt{2}$ (which exceeds the critical impact parameter $\tilde{l}_{\rm crit}(\tilde{Q})$ for all $\tilde{Q} \in [0, 0.5]$).}
			\label{fig:5}
		\end{figure}

		[5]To display the charge dependence directly, we fix the impact parameter $\tilde{l} = 2\sqrt{2}$ (corresponding to $\tilde{r}_m = 2.0$ at $\tilde{Q} = 0$) and scan $\tilde{Q}$ from $0$ to $0.49$, determining $\tilde{r}_m$ for each $\tilde{Q}$ by Eq.~\eqref{eq:lOfRm}. As shown in Figure~\ref{fig:5}, $|\tilde{\vartheta}_\infty|$ decreases monotonically with increasing $\tilde{Q}$.

		\section{Conclusion}
		\label{sec:5}

		In this work, we extend the spinoptics description of the G-SHE to a charged static spherically symmetric spacetime. Using the closed conformal Killing-Yano tensor and its Hodge dual, we construct the parallel-propagated null tetrad along the null geodesic. From this tetrad we compute the driving force $w^\mu$ and give the evolution equations for the rescaled deviation $\xi\equiv r\Psi$ and the tilting angle $\vartheta$.

		The main result is that the G-SHE becomes weaker with increasing charge. This charge dependence is already visible in the evolution equation~\eqref{eq:dxi_tilde}, where the $\tilde{Q}^2$ contribution enters with a sign opposite to the second term of $\left(2\tilde{Q}^2-\frac{3}{2}\tilde{r}\right)$. As the turning point approaches the photon sphere, the magnitude $|\tilde{\vartheta}_\infty|$ grows firstly, but remains bounded, and then exhibits an oscillatory dependence on the location of the turning point when the turning point is very close to the photon sphere; a divergence occurs only on the exactly critical orbit, where the ray remains on the circular photon orbit and the tilting angle grows linearly with the affine parameter (Appendix~\ref{app:critical}). Numerical integration confirms that, at a fixed impact parameter away from the critical value, the dimensionless tilting angle $\tilde{\vartheta}_\infty$, which measures the angular deflection accumulated over the entire scattering process, decreases in magnitude with increasing $\tilde{Q}$. Since $\vartheta_\infty = s\tilde{\lambdabar}\,\tilde{\vartheta}_\infty$, the physical tilting angle exhibits the same charge dependence and remains small whenever the spinoptics approximation is applicable.

		The Kerr--Newman spacetime, which admits a closed conformal Killing-Yano tensor, is a natural next step for extending the present analysis. In particular, it would be interesting to see whether the charge dependence of the G-SHE found here persists when rotation is included.

		\appendix
		\section{Derivation of the polar deviation equation}
		\label{app:deviation}
		
		In this appendix, we verify that the $t$, $r$ and $\phi$ components of $\delta x^\mu$ vanish, and derive Eq.~\eqref{eq:thetaeq} from the deviation equation~\eqref{eq:deviation}. We work on the equatorial plane ($\theta=\pi/2$), keep terms up to first order in $\delta x^\mu$, use the same parameter $\lambda$ for the two rays, and fix the initial data as in~\eqref{eq:IC}.
		
		Equation~\eqref{eq:deviation} is a second-order linear system for $\delta x^\mu$,
		\begin{equation}
			\frac{d^2\delta x^\mu}{d\lambda^2} + \left[ 2\left. \Gamma^\mu_{\ \nu\sigma} \right|_{\gamma(\lambda)} \frac{dx_0^\nu}{d\lambda} \right] \frac{d\delta x^\sigma}{d\lambda} + \left[ \frac{dx_0^\nu}{d\lambda}\frac{dx_0^\sigma}{d\lambda} \left. \partial_\rho\Gamma^\mu_{\ \nu\sigma}(x_0) \right|_{\gamma_0(\lambda)} \right] \delta x^\rho = \left. w^\mu \right|_{\gamma(\lambda)},
			\label{eq:app_system}
		\end{equation}
		whose right-hand side is non-zero only for $\mu=\theta$. The equation for a given $\mu$ can contain $\delta x^\theta$ only through the two terms
		\begin{equation}
			2\Gamma^\mu_{\ \nu\theta}l_0^\nu\, \frac{d\delta x^\theta}{d\lambda}, \qquad l_0^\nu l_0^\sigma\, \partial_\theta\Gamma^\mu_{\ \nu\sigma}\, \delta x^\theta ,
			\label{eq:app_comb}
		\end{equation}
		and for $\mu=t,r,\phi$ neither of them contributes at order $\mathcal{O}(\omega^{-1})$: for $\mu=t$ they vanish identically, since $\Gamma^t_{\ \nu\theta}=0$ and $\partial_\theta\Gamma^t_{\ \nu\sigma}=0$; for $\mu=r$ the first vanishes because $l_0^\theta=0$ and the second because $\partial_\theta\Gamma^r_{\ \nu\sigma}=0$ on the equatorial plane; for $\mu=\phi$ the first is of order $\mathcal{O}(\omega^{-2})$, since $\Gamma^\phi_{\ \phi\theta}|_{\gamma}=\cot\theta=\mathcal{O}(\omega^{-1})$, and the second vanishes because $l_0^\theta=0$. Therefore the $t$, $r$ and $\phi$ equations do not contain $\delta x^\theta$ and form a closed linear homogeneous system.
		
		For $\mu=\theta$ the terms~\eqref{eq:app_comb} are the only ones that survive, and, with $\Gamma^\theta_{\ r\theta}=1/r$, $d\phi/d\lambda=l/r^2$ and $\partial_\theta\Gamma^\theta_{\ \phi\phi}=1$ on the equatorial plane, Eq.~\eqref{eq:deviation} reduces to
		\begin{equation}
			\frac{d^2\vartheta}{d\lambda^2} + \frac{2}{r}\frac{dr}{d\lambda}\frac{d\vartheta}{d\lambda} + \frac{l^2}{r^4}\vartheta = w^\theta ,
		\end{equation}
		which is exactly Eq.~\eqref{eq:thetaeq}.
	
		It remains to show that the $t$, $r$ and $\phi$ components vanish. Their equations form a linear homogeneous second-order system for $\delta x^{t,r,\phi}$, whose only solution with the vanishing initial data~\eqref{eq:IC} is the trivial one. Hence, up to order $\mathcal{O}(\omega^{-1})$, only the polar component survives, as stated in Eq.~\eqref{eq:thetaonly}.

		\section{Behavior on the critical orbit}
		\label{app:critical}
		
		In this appendix, we show that the divergence of the accumulated tilting angle occurs only when the light ray travels on the critical orbit, i.e., the photon sphere, $\tilde{r}_m=\tilde{r}_{\rm ph}$, for which $\tilde{l}=\tilde{l}_{\rm crit}$. On this orbit, Eqs.~\eqref{eq:dxi_tilde} read
		\begin{equation}
			\frac{d\tilde{\xi}}{d\tilde{\lambda}}=\left(2\tilde{Q}^2-\frac{3}{2}\tilde{r}\right)\frac{\tilde{l}_{\rm crit}\tilde{\lambda}}{\tilde{r}^5}
			-\tilde{\vartheta}\,\frac{\tilde{l}_{\rm crit}^2}{\tilde{r}^2},
			\qquad
			\frac{d\tilde{\vartheta}}{d\tilde{\lambda}}=\frac{\tilde{\xi}}{\tilde{r}^2}.
			\label{eq:critSystem}
		\end{equation}
		Differentiating the second equation of the above and substituting the first one into it, we obtain
		\begin{equation}
			\frac{d^2\tilde{\vartheta}}{d\tilde{\lambda}^2}
			+\frac{2}{\tilde{r}}\frac{d\tilde{r}}{d\tilde{\lambda}}\frac{d\tilde{\vartheta}}{d\tilde{\lambda}}
			+\frac{\tilde{l}_{\rm crit}^2}{\tilde{r}^4}\,\tilde{\vartheta}
			=\left(2\tilde{Q}^2-\frac{3}{2}\tilde{r}\right)\frac{\tilde{l}_{\rm crit}\tilde{\lambda}}{\tilde{r}^7}.
			\label{eq:critSecond}
		\end{equation}
		On the circular orbit $\tilde{r}=\tilde{r}_{\rm ph}$, we have $d\tilde{r}/d\tilde{\lambda}=0$, so that
		\begin{equation}
			\frac{d^2\tilde{\vartheta}}{d\tilde{\lambda}^2}+\Omega^2\tilde{\vartheta}=a\,\tilde{\lambda},
			\qquad
			a=\frac{\bigl(2\tilde{Q}^2-\frac{3}{2}\tilde{r}_{\rm ph}\bigr)\tilde{l}_{\rm crit}}{\tilde{r}_{\rm ph}^7},
			\qquad
			\Omega^2=\frac{\tilde{l}_{\rm crit}^2}{\tilde{r}_{\rm ph}^4}.
			\label{eq:critOsc}
		\end{equation}
		This is an undamped harmonic oscillator driven by a force growing linearly with $\tilde{\lambda}$. With the initial conditions $\tilde{\vartheta}(0)=0$ and $\left.d\tilde{\vartheta}/d\tilde{\lambda}\right|_{\tilde{\lambda}=0}=0$ [Eq.~\eqref{eq:numIC}], its solution is
		\begin{equation}
			\tilde{\vartheta}(\tilde{\lambda})=\frac{a}{\Omega^2}\left(\tilde{\lambda}-\frac{\sin\Omega\tilde{\lambda}}{\Omega}\right).
			\label{eq:critSol}
		\end{equation}
		Hence, when the light ray travels on the photon sphere, the accumulated tilting angle diverges linearly and the assumption of a small deviation eventually fails.
		
		\bibliographystyle{JHEP}
		\bibliography{refs}
		
	\end{document}